\documentstyle[12pt]{article}
\newcommand{\ben}{\begin{equation}}
\newcommand{\een}{\end{equation}}
\newcommand{\bea}{\begin{eqnarray}}
\newcommand{\eea}{\end{eqnarray}}
\newcommand{\nn}{\nonumber \\ }
\newcommand{\hf}{\frac{1}{2}}
\newcommand{\eq}{\begin{equation}}
\newcommand{\en}{\end{equation}}
\newcommand{\eqn}{\begin{eqnarray}}
\newcommand{\enn}{\end{eqnarray}}
\newcommand{\CR}{\nonumber \\}
\newcommand{\twt}{\frac{2}{3}}
\newcommand{\hl}{\\ \hline}
\newcommand{\ad}{{\rm ad}}

\newcommand{\BG}{{\bf g}}

\newcommand{\I}{{\rm i}}
\newcommand{\J}{J_{\rm diag}}
\newcommand{\half}{{1\over2}}
\newcommand{\HBG}{\hat{{\bf g}}}
\newcommand{\pa}{\partial}
\newcommand{\rank}{{\rm rank}}

\newcommand{\tr}{{\rm tr}}

\newcommand{\A}{\alpha}
\newcommand{\B}{\beta}
\newcommand{\D}{\delta}
\newcommand{\DE}{\Delta}
\newcommand{\G}{\gamma}

\newcommand{\vep}{\varepsilon}

\newcommand{\T}{\theta}
\newcommand{\LM}{\Lambda}

\newcommand{\NPB}[1]{{\it Nucl. Phys.} {\bf B#1}}
\newcommand{\PLB}[1]{{\it Phys. Lett.} {\bf B#1}}

\newcommand{\CMP}{{\it Comm. Math. Phys.}}
\def\ord#1#2{{#2\over (z-w)^{#1}}}
\def\ordo#1{{#1\over z-w}}

\def\sdim{{\rm sdim{ }}}
\def\WG{\widehat{G}}

\def\TG{\tilde{G}}
\begin{document}
\renewenvironment{thebibliography}[1]
  { \begin{list}{\arabic{enumi}.}
    {\usecounter{enumi} \setlength{\parsep}{0pt}
     \setlength{\itemsep}{3pt} \settowidth{\labelwidth}{#1.}
     \sloppy
    }}{\end{list}}
\begin{titlepage}
\parindent=1.5pc
\begin{flushright}
NBI-HE-93-18 \\
UTHEP-255 \\
July, 1993
\end{flushright}
\vglue 0.6cm
\begin{center}{{\bf EXTENDED SUPERCONFORMAL ALGEBRAS\\
                \vglue 3pt
               AND FREE FIELD REALIZATIONS \\
               \vglue 3pt
               FROM HAMILTONIAN REDUCTION}
\vglue 1.0cm
{KATSUSHI ITO}\\[.2cm]
{\it Institute of Physics, University of Tsukuba}\\
{\it Ibaraki 305, Japan}\\[.4cm]
{JENS OLE MADSEN}\\[.2cm]
{and}\\[.2cm]
{JENS LYNG PETERSEN}\\[.2cm]
{\it Niels Bohr Institute, University of Copenhagen,}\\
{\it Blegdamsvej 17, DK 2100 Copenhagen \O , Denmark}\\[10pt]

\vglue 0.8cm
}
\end{center}
\vglue 0.3cm
\begin{abstract}
We develop the method of the hamiltonian reduction of affine Lie superalgebras
to obtain explicit and general expressions both for the classical
and the quantum  extended superconformal algebras.
By performing the gauge transformation which connects
the diagonal gauge with the Drinfeld-Sokolov gauge and
considering the quantum corrections, we get generic expressions
for the classical and quantum free field realizations of the algebras.

\end{abstract}
\vglue 0.8cm
\end{titlepage}

Recently, there has been a renewal of interest in the non-linearly extended
superconformal algebras, originally introduced by Knizhnik and Bershadsky
\cite{KnBe}.
Two of the present authors have considered the classification of extended
superconformal algebras in the context of
hamiltonian reduction (\cite{DrSo,Be,BeOo,BaFeFoRaWi}) of Lie superalgebras
in \cite{ItMa}, and the present authors have further explored this, both in
the special case of the so called doubly extended $N=4$ algebra
$\tilde{A}_\gamma$ \cite{ItMaPe} and in the general case
\cite{ItMaPe2,ItMaPe3}.
These algebras have also been classified independently by Bowcock and
by Fradkin and Linetsky \cite{BoFrLi} from a purely algebraic approach.

The method of hamiltonian reduction clarifies the integrable structures of
the hierarchy of the relevant soliton equations \cite{DrSo}.
{}From the point of view of conformal field theories, this approach provides
a powerful technique to obtain the free field realizations of the
algebras. These free field realizations are useful for studying the
representation of the algebra and correlation functions.

In the present paper we develop the technique of hamiltonian reduction
to present simple and completely general expressions for the
algebras.
We find the explicit gauge transformation which gives the classical
free field realization. We then obtain a quantum
expression from the requirement of the closure of the algebras.
The difference with the classical expression may be expressed as
a ``renormalization" of certain constants.
Some preliminary results have been reported in ref. \cite{ItMaPe}.

Let $\BG=\BG_{0}\oplus\BG_{1}$ be a rank $n$ basic classical Lie
superalgebra with an even subalgebra $\BG_{0}$ and an odd subspace
$\BG_{1}$. As described in \cite{ItMa,ItMaPe2,ItMaPe3},
we consider the case ${\bf g}_0 = G \oplus A_1$, in which
we use only the $A_1$ subalgebra for the reduction.
The generators of the Lie superalgebra split
into $A_1$ representations with spins $\{s_i\}$, and the conformal
dimensions for a current in the reduced algebra are $\{ 1+s_i \}$.
The fact that $\BG_0 = G \oplus A_1$ implies that the bosonic sector of the
algebra contains
an energy-momentum tensor of conformal dimension 2 and some dimension 1
affine $\WG$ currents.
In the superalgebras listed in table \ref{ta:cis}, the odd
subspace carries spin $\half$ representations of the $A_1$, and therefore
gives rise to fermionic supercurrents of conformal dimension $\frac{3}{2}$
after the reduction.
In the case of the Lie superalgebra $B(1|n)$, where the even subalgebra is of
the form $\BG_0 = C_n \oplus A_1$, the odd subspace carries spin 1
representations of the $A_1$, giving rise to
fermionic currents with integer spin in the reduced algebra. This type of
algebra falls outside of the scope of this paper, and we will
therefore not consider $B(1|n)$ here.

We introduce some notation. We denote by
$\DE=\DE^{0}\cup\DE^{1}$  the set of roots of $\BG$, where
$\DE^{0}$ ($\DE^{1}$) is the set of even (odd) roots.
(The root systems of those Lie superalgebras that we are interested were
given for example in \cite{ItMaPe2}).
Denote  the set of
positive even (odd) roots by $\DE^{0}_{+}$ ($\DE^{1}_{+}$).
The superalgebra $\BG$ has a canonical basis
$\{ E_{\A}, e_{\G}, h^{i} \}$ ($\A\in\DE^{0}$, $\G\in\DE^{1}$,
$i=1, \ldots ,n$), which satisfies (anti-)commutation relations
\eqn
\mbox{[} E_{\A}, E_{\B} \mbox{]}&=&
\left\{ \begin{array}{ll}
N_{\A,\B} E_{\A+\B}, &\quad \mbox{ for $\A$, $\B$, $\A+\B\in \DE^{0}$},\\
\displaystyle{{2\A\cdot h \over \A^{2}}}, &\quad \mbox{ for $\A+\B=0$, }
\end{array}\right.  \CR
\left\{ e_{\G}, e_{\G'} \right\}&=& N_{\G,\G'} E_{\G+\G'}, \quad
\mbox{ for $\G, \G'\in \DE^{1}$, $\G+\G'\in \DE^{0}$},  \CR
\left\{ e_{\G}, e_{-\G} \right\}&=& \G\cdot h ,
\quad \mbox{for $\G\in\DE^{1}_{+}$}, \CR
\mbox{[} e_{\G}, E_{\A} \mbox{]}&=& N_{\G,\A} e_{\G+\A}, \quad
\mbox{ for $\A\in \DE^{0}$, $\G$,$\G+\A\in \DE^{1}$},  \CR
\mbox{[} h^{i}, E_{\A} \mbox{]}&=&\A^{i}E_{\A}, \quad
\mbox{[} h^{i}, e_{\G} \mbox{]}=\G^{i}e_{\G}.
\enn
The even subalgebra $\BG_{0}$ is generated by $\{ E_{\A}, h^{i} \}$.
The odd subspace $\BG_{1}$ is spanned by $\{e_{\G}\}$.
$\BG_{0}$ acts on $\BG_{1}$ as a faithful, completely reducible
representation. The Killing form $( \ ,\ )$ on $\BG$ is defined by
\eq
(E_{\A}, E_{\B})={2\over \A^{2}}\D_{\A+\B,0}, \quad
(e_{\G}, e_{-\G'})=-(e_{-\G}, e_{\G'})=\D_{\G,\G'},  \quad
(h^{i},h^{j})=\D_{i j},
\en
for $\A,\B\in\DE^{0}$, $\G,\G' \in \DE^{1}_{+}$, $i,j=1,\ldots, n$.

In terms of these generators, we can write the expression for the affine
current as
\eq
J(z)=\sum_{\A\in\DE^{0}}{\A^{2}\over 2}J_{\A}(z)E_{\A}
     +\sum_{\G\in\DE^{1}}j_{\G}(z)e_{\G}+\sum_{i=1}^{n}H^{i}(z)h^{i}.
\label{eq:current}
\en
On the space of currents, the Poisson bracket structure is introduced
by the infinitesimal gauge transformations
$\delta J(z)= [ \LM(z), J(z) ]+ k \pa \LM(z)$, where $\LM(z)\in\BG$ and
$k$ is the level of the affine Lie superalgebra $\HBG$.

In the present case, the odd subspace $\BG_{1}$ belongs to the spin $\half$
representation with respect to the even subalgebra $A_{1}$ and
splits into two parts $\BG_{1}=(\BG_{1})_{+\half}\oplus(\BG_{1})_{-\half}$.
By choosing appropriate simple roots for $\BG$, we can take the root
space $\DE^{1}_{+\half}$, which spans $(\BG_{1})_{+\half}$, to be the space of
odd positive roots $\DE^{1}_{+}=\DE^{1}_{+\half}$, and we find the
relation $\DE^{1}_{+}=\DE^{1}_{-} + \T$.

Denote the set of roots of the subalgebra $G$ by $\DE(G)$.
Each odd space $(\BG_{1})_{\pm\half}$ belongs to a
fundamental representation of $G$ of dimension $|\DE^{1}_{+}|$, this
being the number of positive odd roots. If we denote the
representation matrices in the canonical basis by
$t^{\A}_{\G ,\G '},t^i_{\G ,\G '}$
then we have (using the Jacobi identities)
for $\G\in\DE^1_+$ \cite{ItMaPe3}
\bea
t^{\A}_{\G ,\G '} & = & -N_{\A ,-\G }\D_{\G-\A ,\G'},    \nn
t^i_{\G,\G'}      & = & (\G_{\perp})^i\D_{\G,\G'},       \nn
\tr(t^it^j)       & = & c_F \D_{ij},                     \nn
\tr(t^\A t^\B)    & = & c_F \frac{2}{\A^2} \D_{\A+\B,0}, \nn
t^{\A}_{\G,\G'}   & = & -t^{\A}_{\T-\G',\T-\G},          \nn
t^{\A}_{\G,\G'}   & = & t^{-\A}_{\G',\G}.
\label{eq:basis}
\eea
where $\G_{\perp}$ is the projection of $\G\in\DE^{1}_{+}$ on the root space
of $G$ and the constant $c_{F}$ is
defined as $\sum_{\G\in\DE^{1}_{+}}\G_{\perp}^{2}/\rank (G)$.
It is convenient to introduce the affine $\WG$ currents
in the above basis (\ref{eq:basis}):
\ben
J_{\G,\G'}(z)\equiv\sum_{\A\in\Delta(G)}\frac{\A^2}{2}t^{\A}_{\G,\G'}
J_{-\A}(z)+\sum_{i=1}^{\rank (G)}t^i_{\G,\G'}H_i(z)
\label{eq:JGGp}
\een

The procedure of hamiltonian reduction is to impose the condition
$J_{-\theta}(z)=1$ on the current, where $-\T$ is the negative root in the
$A_1$ subalgebra. With this constraint,
there is still a gauge invariance, which allows one to choose one or the other
gauge slice as a representative of the reduced phase space, for example the
so-called \lq\lq Drinfeld-Sokolov (DS)" gauge or the diagonal gauge.
In the DS gauge the current is
\eq
J_{DS}(z)= \frac{\T^2}{2} E_{-\T} + \frac{\T^2}{2}T(z)E_{\theta}
          +\sum_{\G\in \DE^{1}_{+}}G_{\G}(z)e_{\G}
          +\sum_{\A\in\DE(G)}\frac{\A^2}{2} J_{\A}(z) E_{\A}
          +\sum_{i=1}^{\rank (G)}H^{i}h^{i}.
\label{eq:dsgauge}
\en
The generic gauge transformation projected on the DS gauge slice
becomes the transformation corresponding to the extended superconformal
algebra, see \cite{ItMa,ItMaPe2} for details.
the Poisson bracket structure on the reduced phase space can be conveniently
written in terms of formal ``operator product expansions".
We define the total energy-momentum tensor to be
$T_{ESA} = \frac{\T^2}{2k}T + T_{Sugawara}$, where $T$ is the generator
occurring in the DS current, and
\bea
T_{Sugawara}&=&{1\over 2k}\left\{
\sum_{\A\in\DE(G)} {\A^{2}\over 2}J_{\A}J_{-\A}
     +\sum_{i=1}^{\rank (G)}H^{i}H^{i} \right\}\nn
&=&\frac{1}{2kc_F}tr(J^2),
\eea
and we rescale the supercurrents as
$\TG_\G = \I\sqrt{\frac{\T^2}{2k}} G_\G$.
The result is the classical extended superconformal algebra:
\eqn
T_{ESA}(z)T_{ESA}(w)&=&\ord{4}{{-6k\over \T^{2}}}
         +\ord{2}{2T_{ESA}(w)}
         +\ordo{\pa T_{ESA}(w)}+\cdots, \CR
T_{ESA}(z)\TG_{\G}(w)&=&\ord{2}{{3\over2}\TG_{\G}(w)}+\ordo{\pa\TG_{\G}(w)}
+\cdots, \CR
J_{\B}(z)\TG_{\G}(w)&=&\ordo{t^\B_{\G,\G^\prime}\TG_{\G^\prime}(w)}+\cdots,
\quad
H^{i}(z)\TG_{\G}(w)=\ordo{t^i_{\G,\G^\prime}\TG_{\G^\prime}(w)}+\cdots, \CR
\TG_{\T-\G'}(z)\TG_{\G}(w)&=&-\ord{3}{2 k\D_{\G',\G}}
+\left ( 2 \ord{2}{J_{\G,\G'}(w)} + \ordo{\pa J_{\G,\G'}(w)} \right )\cr
& + & \frac{\T^2}{2}\D_{\G ,\G'}\ordo{T_{ESA}(w)-T_{Sugawara}(w)}
- \frac{1}{k}\ordo{(J^2)_{\G,\G'}(w)}
+\cdots,
\label{eq:clope}
\enn
Here we imply the notation:
\ben
(J^2)_{\G,\G'}\equiv\sum_{\G''\in\DE^1_+}J_{\G,\G''}J_{\G'',\G}
\een
Notice that this expression is nonvanishing only if either $\G-\G'\in\DE(G)$,
$\G =\G'$ or $\G+\G'=\T$.
We note that the classical value of the central charge $c_{ESA}$ is
$-12k/\T^{2}$. The operator product expansions of the affine $\WG$
(or $\WG=\oplus_i\WG^{i}$) currents are the
standard ones at level $K^{(i)}=\frac{2k}{\A_{Li}^2}$, where $\A_L$
denotes the ``long root".

There are several methods which allow us to obtain the classical free field
realization of the extended Virasoro algebras. One method is based on the Miura
transformation, used in the case of hamiltonian reduction for
non-exceptional simple Lie algebras \cite{DrSo}.
This formulation has been generalized to the extended superconformal algebras
for Lie superalgebras $sl(N|2)$ and $osp(N|2)$ \cite{ItMa}.
However for the remaining Lie (super)algebras, especially exceptional type
algebras, it is difficult to find the explicit form of the Miura
transformations.
Another method consists in finding the explicit gauge transformation from the
diagonal gauge, in which the Poisson structure
of the free fields are defined, to the Drinfeld-Sokolov gauge.
We show that this last method provides a general procedure for
giving the free field realization for arbitrary extended algebras.

Let us consider the gauge transformation
$J \rightarrow J^g = gJg^{-1} + k\pa g g^{-1}$.
By parametrizing a supergroup element $g$ by $e^\LM$ ($\LM \in \BG$),
we get
\eq
J^g=\sum_{n=0}^{\infty}{1\over n!} \{ (\ad \LM)^n J
+ k (\ad \LM)^{n-1} \pa \LM\}.
\label{eq:gautr}
\en
Here $(\ad X)Y= [ X , Y ]$ for $X,Y \in \BG$.
We write the current in the diagonal gauge as
\ben
\J(z) = \sum_{\A\in\DE(G)} \frac{\A^2}{2} \hat{J}_{\A}(z) E_{\A}
+ \sum_{i} \hat{H}_{i}(z) h^i
+ \sum_{\G\in\DE^1_+} \I \sqrt{\frac{\T^2}{2}} \chi_{\T-\G}(z) e_{-\G}
- \I\sqrt{k}\pa\phi(z) h_{\T} + \frac{\T^2}{2}E_{-\T},
\label{eq:diaggauge}
\een
where $\chi_\G$ are complex fermions, normalized to
$$
\chi_\G(z)\chi_{\T-\G^\prime}(w) = \frac{\D_{\G,\G^\prime}}{z-w} + \cdots,
$$
and $\phi$ is a free boson coupled to a background charge
$q=-2\sqrt{\frac{k}{\T^2}}$ with
$\phi(z)\phi(w)=-\ln (z-w)+\cdots $.
The energy-momentum
tensor corresponding to $\phi$ is
$T_\phi = -\half(\pa\phi)^2 - \frac{iq}{2}\pa^2\phi$. The currents $\hat{J}$
are affine $\hat{G}$ currents that commute with the fermions. The level of the
$\hat{J}$'s are $K = \frac{2}{\A_L^2}k$. The
$\hat{J}$'s can be built based on the standard Wakimoto
construction \cite{FeFr,ItKo}, but the explicit free field realization of the
currents is not necessary in the present paper.
Using various equations for the structure constants of the Lie superalgebra
\cite{ItMaPe2}, it is straightforward to calculate $(J_{diag})^g$
with $g=e^{\LM}$
and
\eq
\LM(z) = \sum_{\G\in\DE^{1}_{+}}
     \xi_\G(z) e_\G + \vep(z) E_{\T},
\en
which generates the residual gauge symmetry.
The gauge transformation (\ref{eq:gautr})
truncates at fourth order for $J=J_{diag}$. The truncation at finite order is
a general feature of the technique of hamiltonian reduction.
The condition
\eq
J_{DS}=(J_{diag})^{g},
\en
determines $\LM$ to be
$\LM = - \I\sqrt{\frac{2}{\T^2}} \sum_\G \chi_{\T-\G} e_\G
+ \I\sqrt{\frac{k}{\T^2}}\pa\phi E_\T$
and we find
\bea
J_{DS} & = & \I \sqrt{\frac{2}{\T^2}}
\sum_{\G\in\DE^{1}_{+}} \left \{ -k\pa\chi_{\G}
- \frac{\I}{\sqrt{2}}\sqrt{\frac{k\T^2}{2}}\pa\phi\chi_{\G}
- \hat{J}_{\G,\G'}\chi_{\G'} - \frac{2}{3} J^f_{\G,\G'}\chi_{\G'}
\right \} e_{\T-\G} \CR
& & + \sum_{\A} \frac{\A^2}{2} (\hat{J}_{\A} + J^f_{\A}) E_\A
+ \sum_{i} (\hat{H}_{i} + H^f_{i}) h^i \label{eq:ffrep} \\
& & + k \left ( - \half (\pa\phi)^2
+ \I\sqrt{\frac{k}{\T^2}}\pa^2\phi
- \frac{1}{2kc_F} {\rm tr} (J^f)^2
- \frac{1}{kc_F} {\rm tr} (\hat{J}J^f)
+ \sum_{\G\in\DE^{1}_{+}} \pa\chi_{\T-\G}\chi_{\G} \right ) E_\T.\nonumber
\eea
Comparing (\ref{eq:dsgauge}) and (\ref{eq:ffrep}) we get the
free field representation of the classical extended superconformal
algebras.
The energy momentum tensor and the supercurrents are given by
\bea
T_{ESA}
& = & -\half(\pa\phi)^2 + \I\sqrt{\frac{k}{\T^2}}\pa^2\phi + \hat{T}_{Sugawara}
+ \half \sum_{\G\in\DE^1_+} \pa\chi_{\T-\G}\chi_\G   \cr
\tilde{G}_\G & = & + \sqrt{k}\pa\chi_{\G}
+ \frac{\I}{\sqrt{2}}\sqrt{\frac{\T^2}{2}}\pa\phi\chi_{\G}
- \frac{1}{\sqrt{k}} \sum_{\G'\in\DE^1_+}
\left ( \hat{J}_{\G,\G'} + \frac{2}{3} J^f_{\G,\G'}
\right ) \chi_{\G'}.
\label{eq:clff}
\eea
The free field realizations of the affine currents $J_\A$ and $H^i$ are
just the coefficients of $\frac{\A^2}{2}E_{\A}$ and $h^i$ respectively :
\bea
J_\A & = & \hat{J}_{\A} + J^f_{\A},  \cr
H^i  & = & \hat{H}_{i} + H^f_{i},
\label{eq:JH}
\eea
where $J^f_{\A}$ and $H^f_i$ are Kac-Moody currents constructed from the
fermions in the standard way
\bea
J^f_{\A} & = &
\half\sum_{\G',\G \in \DE^1_+}t^\A_{\G',\G}\chi_{\G}\chi_{\T-\G'}  \cr
H^f_{i} & = &
\half\sum_{\G',\G \in \DE^1_+}t^i_{\G',\G}\chi_{\G}\chi_{\T-\G'}.
\eea

We now use the classical free field realizations
to write down an ansatz for the generators of the quantum algebra algebra,
in which we allow for arbitrary coefficients on the various terms.
We then calculate the quantum operator product expansions of these generators.
This way we find the coefficients and the modifications in the algebra
and the free field
realizations necessary for the closure of the quantum algebra.
Define $\chi_\G$ and $\hat{J}$ as above, and define $\phi$ to be coupled to
the background charge
$Q=-\frac{2}{\sqrt{\T^2}\A_+}(k+\frac{\T^2}{2})$, where
$\A_+ = \sqrt{k+h^\vee}$.
$h^{\vee}$ is the dual Coxeter number of the Lie superalgebra $\BG$.
The affine currents $J_\A$ and $H^i$ with level
$K = \frac{2}{\A_L^2}(k-\frac{c_F}{2})$ are constructed as in eq. (\ref{eq:JH})
>from the affine currents $\hat{J}_\A$ and $\hat{H}^i$ from the free field
representation, with level
$\hat{K}=\frac{2}{\A_{L}^2}(k-c_F)$, and the affine currents $J^f_\A$ and
$H^i_f$
constructed from the fermions, with level $K^f = \frac{c_F}{\A_L^2}$.
{}For the explicit realizations of the affine currents $\hat{J}_\A$ and
$\hat{H}^i$ in terms of free fields
we refer to ref. \cite{FeFr} or \cite{ItKo}.
The quantum free field realizations of the supercurrents are
found to be
\ben
G_{\G}(z) =
\frac{(k+\theta^2/2)}{\A_+}\pa\chi_{\G}
+ \frac{i}{\sqrt{2}}\sqrt{\frac{\theta^2}{2}}\pa\phi(z)\chi_{\G}
- \frac{1}{\A_+}\sum_{\G'\in\DE^1_+}
\left ( \hat{J}_{\G,\G'}(z)\chi_{\G'}(z)
+ \frac{2}{3} :J^{f}_{\G,\G'}\chi_{\G'}:(z) \right )
\label{eq:qffG}
\een
The quantum energy-momentum tensor is
$T_{ESA}(z) = T_\phi(z) + T_\chi(z) + \hat{T}_{Sugawara}(z)$, where
\bea
T_\phi(z) & = & -\half:(\pa\phi)^2:(z) - \frac{iQ}{2}\pa^2\phi(z),    \cr
\hat{T}_{Sugawara}(z) & = & {1\over 2\A_+^2}\left\{
\sum_{\A\in\DE(G)} {\A^{2}\over 2}:\hat{J}_{\A}\hat{J}_{-\A}:(z)
     +\sum_{i=1}^{\rank (G)}:\hat{H}^{i}\hat{H}^{i}:(z) \right\}, \cr
T_\chi(z) & = & \half \sum_{\G\in\DE^1_+} :\pa\chi_{\T-\G}\chi_\G:(z).
\label{eq:qffT}
\eea
The quantum algebra is very similar to the classical one. The central charge of
the quantum algebra is shifted from $-\frac{12k}{\T^2}$ to
$-\frac{12k}{\T^2} + \frac{k\cdot\sdim \BG}{k+h^\vee} - 2 -\half |\DE^1_+|$
where
$\sdim \BG$ is the superdimension of $\BG$, and as already mentioned above, the
level
of the affine currents is shifted from $K_{cl} = \frac{2}{\A_L^2}k$ to
$K = \frac{2}{\A_L^2}(k-\frac{c_F}{2})$. The OPE between
the supercurrents are modified into
\bea
G_{\T-\G'}(z)G_{\G}(w)
&=&\frac{-f(k)}{k+h^{\vee}}
\frac{\D_{\G',\G}}{(z-w)^3}
+\frac{1}{k+h^{\vee}} \sum_i g^{(i)}(k) \left (
2\frac{J^{(i)}_{\G,\G'}(w)}{(z-w)^2}
+\frac{\pa J^{(i)}_{\G,\G'}(w)}{z-w}
\right )\nn
&+&\!\!\!\!
\frac{\T^2}{2}\D_{\G,\G'}\frac{T_{ESA}(w)-T_G(w)}{z-w}
-\frac{1}{k+h^{\vee}} \sum_i \frac{:(J^2)_{\G,\G'}:_S(w)}{z-w}+\cdots .
\label{eq:GG}
\eea
Here $J^{(i)}_{\G,\G'}$ denotes the part of the current $J_{\G,\G'}$ associated
with the affine algebra $\WG^{(i)}$. The functions $f(k)$ and $g(k)$ are given
by
\bea
f(k) & = & 2(k+\frac{\T^2}{2}+\frac{\A_L^2}{4}H^\vee)(k-\frac{c_F}{2}), \cr
g^{(i)}(k) & = & 2(k+\frac{\T^2}{2}+\frac{(\A^{(i)}_L)^2}{4}(H^\vee)^{(i)}),
\eea
In this expression it is not obvious that $f(k)$ is well defined in the case
where $G = G^{(1)} \oplus G^{(2)}$. However, one finds the identity
$$
\frac{\T^2}{2} + \left (\frac{\A_L^2}{4}H^\vee \right )^{(1)}
= - \frac{c_F^{(2)}}{2},
$$
and similarly when (1) and (2) is interchanged. One easily checks that this
does indeed imply that $f(k)$ is well defined.

We note that in the classical limit $k\rightarrow \infty$ we recover the
classical expression (\ref{eq:clope}).

$$T_G(w)\equiv\frac{1}{2(k+h^{\vee})}\{\sum_{\A\in\DE(G)}\frac{\A^2}{2}
:J_{\A}J_{-\A}:(w)+\sum_i:H_iH_i:(w)\}$$
is similar to the Sugawara energy momentum tensor, but in fact carries a
different normalization, namely the normalization that would be needed for
the ``hat" part whereas here we are dealing with the full generators.
In order to present the algebra in the above way which is related so simply
to the classical expressions,
it is important to use the {\em symmetrized} normal ordering prescription
in the quadratic terms in the expression for the ope
$G_{\T-\G'}(z)G_{\G}(w)$ :
$$
:A(z)B(z):_S \equiv \hf\oint_z\frac{dw}{2\pi i}\frac{A(z)B(w)+A(w)B(z)}{z-w}.
$$

In this paper we have employed a unified formalism for treating those
superconformal algebras which
are obtainable by hamiltonian reduction of Lie superalgebras, using the
techniques developed in refs. \cite{ItMa,ItMaPe,ItMaPe2}. The new results
of the present paper are that we have obtained
expressions for the algebras and for the free field realizations of these
algebras, which are not only completely general, but also very simple compared
to previous expressions.
We have provided formulas both for the classical
(\ref{eq:clope},\ref{eq:clff}) and the
quantum (\ref{eq:qffG},\ref{eq:qffT},\ref{eq:GG}) cases.
The two are quite similar but differ as expected by certain ``renormalizations"
of constants for which we have obtained the general form depending on certain
group theory parameters.
As a particular new result in this paper we have obtained the classical
expressions for the free field realizations in a very straight forward way,
simply by providing an
explicit expression for the gauge transformation connecting the
diagonal gauge (\ref{eq:diaggauge}) and the Drinfeld-Sokolov gauge
(\ref{eq:ffrep}). This method would seem to be
useful for finding the free field realization for other extended algebras.

The study of conformal field theories with extended chiral algebras containing
a superconformal algebra, is expected to be of interest from various points of
view. Thus the corresponding target manifold in a string
theory application would be expected to have a rich geometrical structure.
{}For actual applications we need more information on the representation theory
of these algebras. Some crucial steps towards an understanding of the
representations were taken
in \cite{ItMaPe,ItMaPe2} where we considered the structure
of the screening operators and the null vectors of these algebras.
But more work in this direction remains. Furthermore it can be expected that
for actual applications the relevant chiral algebra will contain one of
the superconformal algebras described here
(i.e. with generators of dimensions $1, 3/2$ and 2, only)
as a subalgebra. This points to the
much larger study of super-$W$-algebras and their classification. A
particularly interesting relation should exist between the algebras we have
studied  here and the ones considered in
\cite{FrRaSo} where the extended superconformal algebras obtainable from
a similar but supersymmetric hamiltonian reduction have been classified.
Indeed the two classifications agree in many ways.
However that
formalism results in algebras with generators appearing as supersymmetry
multiplets.
The relation between that formalism, and the one we use in this paper
has recently been considered in \cite{Ra}.

\vglue 0.6cm
{\bf Acknowledgements:} This work was supported in part by EEC contract
no. SC1 0394 C (EDB).
One of the authors (K.I.) would like to thank the Niels Bohr Institute for
its financial support during his stay.
The free field realizations in this paper has been checked using the
Mathematica operator product expansion package constructed by Kris Thielemans.
\vglue 0.6cm
\vglue 0.6cm
\begin{table}
\caption{Constants occurring in the free field realizations.}
\label{ta:cis}
\begin{center}
$$
\begin{array}{|l|lrrrrrr|}                         \hline\hline
\BG   &  G   & \theta^2/2 & h^\vee & (\frac{\A_L^2}{2}H^{\vee})^{(1)}
& (\frac{\A_L^2}{2}H^{\vee})^{(2)} & c_F^{(1)} & c_F^{(2)}
\hl
B(n|1)     &            B_n &   2 & -2n+3 & -2n+1 &         &  -2 &          \\
D(n|1)     &            D_n &   2 & -2n+4 & -2n+2 &         &  -2 &          \\
{}F(4)     &            B_3 &\twt &    -3 &    -5 &         &  -2 &          \\
G(3)       &            G_2 &-\frac{4}{3} & 2 & 4 & & 2 &                    \\
A(n|1)     & A_n\oplus u(1) &  -1 &   n-1 &   n+1 &       0 &   2 &   -n+1 \\
D(2|n)     & C_n \oplus A_1 &  -1 &  2n-2 &  2n+2 &      -2 &   4 &   -2n \\
D(2|1;\A)  & A_1 \oplus A_1 &   1 &     0 &  -2\G &-2(1-\G) &-2\G &
 -2(1-\G) \hl
\end{array}
$$
\end{center}
\end{table}

\newpage
\vspace{.5cm}
{\bf References}
\vglue 0.3cm


\begin{thebibliography}{99}
%
\bibitem{KnBe}
V.G.~Knizhnik, Theor.~Math.~Phys. {\bf 66} (1986) 68;
M.~Bershadsky, \PLB{174} (1986) 285.
\bibitem{DrSo}
V.G.~Drinfeld and V.V.~Sokolov, J. Sov. Math. {\bf 30} (1984) 1975.
\bibitem{Be}
A.A.~Belavin, in {\it ``Quantum String Theory"
Proceedings in Physics vol.} {\bf 31} (Springer Verlag, Berlin ,1989);
A.~Alekseev and S.~Shatashvili, \NPB{323} (1989) 719.
\bibitem{BeOo}
M.~Bershadsky and H.~Ooguri, Comm. Math. Phys. {\bf 126} (1989) 429.
\bibitem{BaFeFoRaWi}
J.~Balog, L.~Feh\'er, P.~Forg\'acs, L.~O'Raifeartaigh and A.~Wipf,
Ann. Phys. { 203} (1990) 76.
\bibitem{ItMa}
K.~Ito and J.O.~Madsen, \PLB{283} (1992) 223.
\bibitem{ItMaPe}
K.~Ito, J.O.~Madsen and J.L.~Petersen,
{\it Phys. Letters  } {\bf B292} (1992) 298.
\bibitem{ItMaPe2}
K.~Ito, J.O.~Madsen and J.L.~Petersen, {\it Nucl. Phys. }{\bf B398} (1993) 425.
\bibitem{ItMaPe3}
K.~Ito, J.O.~Madsen and J.L.~Petersen, {\it NBI preprint NBI-HE-92-81;}
to appear in the proceedings of the International Workshop on "String Theory,
Quantum Gravity and the Unification of the Fundamental Interactions", Rome,
September 21-26 1992.
\bibitem{BoFrLi}
P.~Bowcock, {\it Nucl. Phys.} {\bf B381} (1992) 415.
E.S.~Fradkin and V.Ya.~Linetskii, {\it preprint ITP-SB-92-34},
{\it preprint ETH-TH/92-6.}
\bibitem{FeFr}
M.~Wakimoto,  \CMP {\bf 104} (1986) 605;
D.~Bernard and G.~Felder, \CMP {\bf 127} (1990) 145;
B.~Feigin and E.~Frenkel, in  {\it ``Physics and Mathematics of
Strings",} eds.  L.~Brink et al., (World Scientific, 1990
Singapore).
\bibitem{ItKo} K.~Ito and S.~Komata, Mod. Phys. Lett. {\bf A 6} (1991) 581.
\bibitem{FrRaSo} L. Frappat, E. Ragoucy and P. Sorba, {\it preprint
ENSLAPP-AL-391/92};
\bibitem{Ra} E. Ragoucy, {\it preprint NORDITA-93-39-P};
\end{thebibliography}
\end{document}